\documentclass[prb,twocolumn,showpacs,preprintnumbers,amsmath,amssymb]{revtex4-1}

\usepackage{graphicx}
\usepackage{dcolumn}
\usepackage{bm}
\usepackage{psfrag}
\usepackage{epsfig}
\usepackage{amsmath}
\usepackage{amssymb}
\usepackage{color}
\usepackage{bbm}
\usepackage[FIGTOPCAP,raggedright,nooneline]{subfigure}

\newcommand{\ignore}[1]{}

\begin{document}

\title{Phonon Cooling and Lasing with Nitrogen-Vacancy Centers in Diamond
}

\author{K. V. Kepesidis$^1$, S. D. Bennett$^2$, S. Portolan$^1$, M. D. Lukin$^2$, and P. Rabl$^1$} 

\affiliation{$^1$Institute of Atomic and Subatomic Physics, TU Wien, Stadionallee 2, 1020 Wien, Austria}
\affiliation{$^2$Physics Department, Harvard University, Cambridge, Massachusetts 02138, USA}

\date{\today}

\begin{abstract}
We investigate the strain-induced coupling between a nitrogen-vacancy impurity and a resonant vibrational mode of a diamond nanoresonator. We show that under near-resonant laser excitation of the electronic states of the impurity, this coupling can modify the state of the resonator and either cool the resonator close to the vibrational ground state or drive it into a large amplitude coherent state. We derive a semi-classical model to describe both effects and evaluate the stationary state of the resonator mode under various driving conditions. In particular, we find that by exploiting resonant single and multi-phonon transitions between near-degenerate electronic states, the coupling to high-frequency vibrational modes can be significantly enhanced and dominate over the intrinsic mechanical dissipation. Our results show that a single nitrogen-vacancy impurity can provide a versatile tool to manipulate and probe individual phonon modes in nanoscale diamond structures.  
 \end{abstract}

\maketitle

%
%

Diamond has emerged as a promising material
for quantum applications, due in part to its
optical and mechanical properties and in part to its
addressable quantum defects. The most widely studied defect is
the negatively charged nitrogen-vacancy (NV) center~\cite{NVReview1,NVReview2},
whose electronic spin exhibits exceptionally long coherence times~\cite{spin-coherence}
and can be prepared and detected optically~\cite{JelezkoPRL2004}.
It has been demonstrated that the NV electronic spin can be entangled with and via optical photons \cite{NatureLukin,BernienNature2013},
and significant effort has been devoted to fabricating nanophotonic
structures to create enhanced NV-photon interfaces \cite{Babinec2011,BaynNJP2011,RiedrichNatNano,FaraonPRL2012}
for efficient quantum information processing and quantum communication.
In parallel, diamond nanostructures
can be fabricated with very high mechanical
quality factors\cite{Ovartchaiyapong2012,Tao2012}, and it has been
proposed theoretically to exploit
the coupling of NV centers to phonons, in addition to photons,
for quantum information processing \cite{RablNatPhys2010,HabrakenNJP2012,Albrecht2013}
or quantum enhanced magnetometry \cite{spin-spin} applications.

It is well known that many electronic defects in solids, including NV centers,
are highly susceptible to deformations of the surrounding lattice.
One consequence
is the phonon-induced broadening of optical lines.
Recently, there has been significant interest in 
exploiting 
these defect-phonon interactions in nanomechanical 
systems or phonon cavities, where 
single defects may be strongly coupled to
long-lived, spectrally-isolated
 phonon modes~\cite{RemusPRB2009,SoykalPRL2011,Ruskov2012,RamosPRL2013,HabrakenNJP2012,Albrecht2013,spin-spin,BurekNanoLett2012,HausmannPSSA2012}.  
This suggests a route toward cavity quantum
electrodynamics using phonons, with applications ranging 
from measurement and manipulation of single mechanical quanta,
to the generation of single-phonon nonlinearities 
and phonon-meditated coupling of defects.
In view of recent advances in diamond nanofabrication 
and demonstrated optical control of NV centers, 
diamond is a leading candidate material 
in which to pursue these directions in experiments.

In this paper, we consider the strain coupling between a 
single NV center and a single resonant
mechanical mode of a diamond nanoresonator, 
and analyze ground state cooling~\cite{WilsonRaePRL2004,MartinPRB2004,ZhangPRL2005,JaehneNJP2009,RablPRB2009,ZippilliPRL2009,prabl} and phonon lasing~\cite{Bennett, Micromaser,Lasing2,a-phonon-laser,phonon-laser-action,Lasing,quantum-dots,phonon-lasing} techniques for manipulating the phonon mode in this system. 
Compared to previous proposals for using the strain coupling to natural and artificial two level defects~\cite{WilsonRaePRL2004,quantum-dots,Yeo} to achieve this task,
%
we here exploit the rich electronic structure of the NV center
and focus on an approach involving two near-degenerate electronically excited states. 
The presence of this additional third defect state leads to qualitatively new features, 
and can be used to resonantly enhance defect-phonon interactions.
These resonances can significantly increase
both cooling and lasing, and are especially important when the
phonon frequency is high---as is the case in small diamond
resonators---in which case 
the standard off-resonant approach is inefficient.
Our results have direct implications for
ground state cooling and quantum
state preparation of phonons in diamond nanoresonators.
More importantly, our approach to phonon lasing enables
a new method for local actuation of
high frequency acoustic modes, providing a useful
tool to measure, control, and characterize
NV-phonon interactions in nanoscale structures.

The paper is structured as follows. In Sec.~\ref{sec:Idea} we start with a brief outline of the basic ideas and main findings of this work. In Sec.~\ref{sec:Model} we present a more detailed derivation of the effective model for describing the NV-phonon coupling, which we then use in Sec.~\ref{sec:Cooling} and Sec.~\ref{sec:Lasing} to study cooling and lasing effects in the off-resonant and resonant regime. 
Finally, in Sec.~\ref{sec:Spectrum} we discuss signatures of cooling and lasing effects in the excitation spectrum of the NV center and summarize the main results and conclusions of this work in Sec.~\ref{sec:Conclusions}.

\section{Idea and Approach}\label{sec:Idea}

\begin{figure}
 \begin{center}
\includegraphics[width=0.48\textwidth]{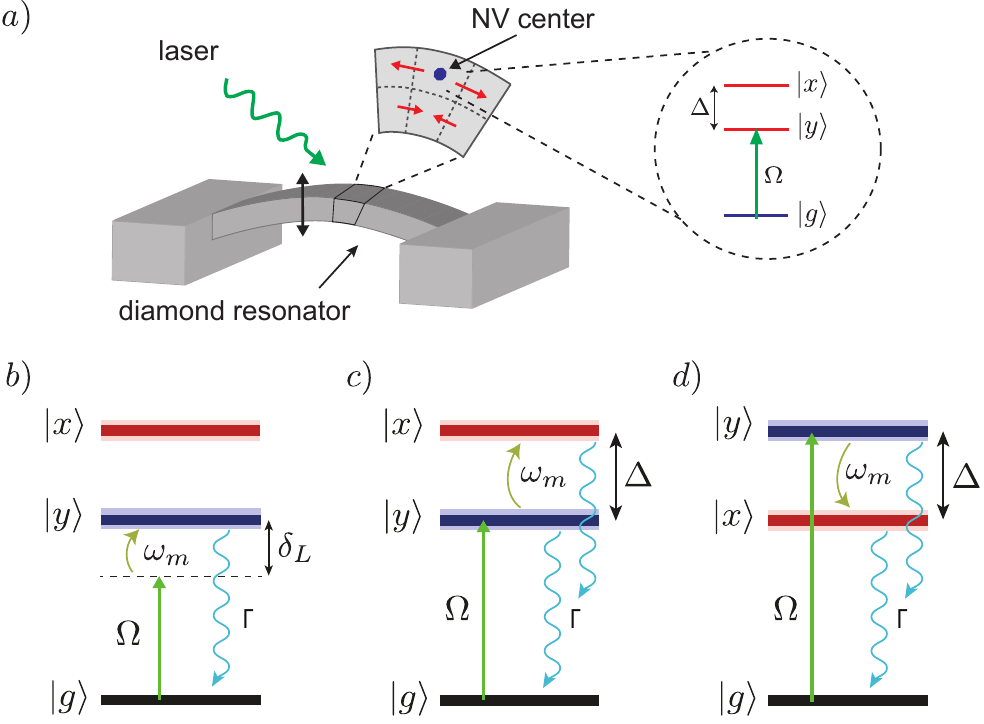}
  \caption{(color online). a) Setup. A single NV$^-$ defect center is embedded in an all-diamond doubly clamped beam. Vibrations of the beam with frequency $\omega_m$ modulate the local strain and shift the energy levels of the electronically excited defect states. b)-d) Illustration of the phonon-assisted optical transitions for the case where the state $|y\rangle$ is driven by a laser of frequency $\omega_L$ and detuning $ \delta_L = \omega_L - \omega_y $. The spacing between the two excited levels is defined as $ \Delta = \omega_x - \omega_y $. 
  b) Phonons  coupled to $\Sigma_\parallel$ only affect the driven state $|y\rangle$ and lead to cooling (heating) effects for  a laser detuning $ \delta_L \approx -\omega_m$ ($ \delta_L \approx +\omega_m $). c) For phonon-induced transitions between the excited states with
  coupling $\sim \Sigma_\perp$, a resonant cooling process occurs for $ \Delta = \omega_m$ and with resonant excitation of the $|y\rangle$ state.  d) For an opposite level ordering, i.e.  $\Delta = -\omega_m$, the same process leads to resonant phonon emission, 
 leading to heating and lasing effects discussed in
Sec.~\ref{sec:Lasing}. }
 \label{fig:Setup}
 \end{center}
 \end{figure}

The basic idea of this work is illustrated by the schematic setup shown in Fig.~\ref{fig:Setup} a), where a single NV center is embedded in  
a diamond nanobeam or other vibrating structure. 
The negatively charged NV$^-$ center in diamond is formed by a substitutional nitrogen atom and an adjacent lattice vacancy; by ignoring spin degrees of freedom for the moment, the electronic level structure of this defect is well described by a single electronic ground state $|g\rangle$ and two optically excited states $|x\rangle$ and $|y\rangle$ [see Sec.~\ref{sec:Model} for a more detailed discussion].  
Due to the $C_{3v}$ symmetry of the NV center, the states $|x\rangle$ and $|y\rangle$ are degenerate in energy, but can be split by a few GHz in the presence of static lattice distortions or by applying external electric fields. 
At cryogenic temperatures, the linewidth of the excited states is sufficiently narrow such that they can be selectively addressed by laser fields of appropriate linear polarization\cite{NatureLukin, PRL2009}.

\subsection{NV-phonon interaction}

The degeneracy of the excited $|x\rangle$ and $|y\rangle$ orbitals makes these states highly susceptible to variation of the local strain near an NV center.
Here, we are interested in the resulting coupling of the NV center to the quantized strain field associated with a single resonant vibrational mode of a diamond structure. 
In general, the strain field induced by this mode will break the symmetry of the NV center and cause energy shifts as well as a mixing of the states $|x\rangle$ and $|y\rangle$.   
The resulting NV-phonon coupling is of the form $(\hbar=1)$
\begin{equation} \label{eq:HNVphon} 
H_{\rm NV-ph}\simeq \left(\lambda_\parallel \Sigma_\parallel + \lambda_\perp\Sigma_\perp\right)  (a^\dag + a), 
\end{equation} 
where $a$ and $a^\dag$ are the annihilation and creation operators for the vibrational mode and $\Sigma_\parallel=|x\rangle\langle x| -|y\rangle\langle y|$ and $ \Sigma_\perp=|x\rangle\langle y| +|y\rangle\langle x|$ are the operators associated with a relative energy shift and a mixing between the excited states, respectively.
For beam dimensions on the scale of $\sim \mu$m, the lowest vibrational modes have mechanical frequencies in the range of $\omega_m\sim 0.1-10$ GHz and the coupling constants $\lambda_{\parallel}$ and $\lambda_{\perp}$ can reach values of several MHz. This is comparable to the radiative lifetime $\Gamma$ of the excited states and can be even stronger in smaller structures~\cite{SoykalPRL2011,Albrecht2013}. More importantly for the present work, the strength of the NV phonon coupling  can by far exceed the mechanical damping rate $\gamma_m=\omega_m/Q$, which for realistic mechanical quality factors of $Q=10^5-10^6$ is in the kHz regime.


\subsection{Phonon cooling and lasing in the resonant and off-resonant regime}\label{sec:cooling-lasing}

The strain coupling given in Eq.~\eqref{eq:HNVphon} describes modulations of the NV excited state level configuration by the mechanical mode. Under laser excitation this gives rise to additional phonon-assisted processes depicted in Fig.~\ref{fig:Setup} b)-d), which depending on the choice of the laser detuning, reduce  (phonon absorption) or increase (phonon emission) the mechanical energy. If the rate $\tilde \Gamma$ associated with these processes substantially exceeds the intrinsic mechanical damping rate $\gamma_m$, the mechanical mode can be cooled close to the quantum ground state.
On the other hand, in the opposite regime, the mechanical mode can
be actuated and driven into a large amplitude coherent state (`phonon lasing').  

In Sec.~\ref{sec:Cooling} and Sec.~\ref{sec:Lasing} we discuss in detail the cooling and lasing effects in this system as a function of the driving laser parameters. From this analysis we find a significant quantitative difference for processes related to the $\Sigma_\parallel$ and $\Sigma_\perp$ type couplings appearing in Eq.~\eqref{eq:HNVphon}. The first case $\sim \Sigma_\parallel(a+a^\dag)$ represents an off-resonant interaction, where only the energy of the driven excited state is modulated. This situation is similar to the coupling of nanomechanical systems to quantum dots or other solid state two level systems, where cooling~\cite{WilsonRaePRL2004,MartinPRB2004,ZhangPRL2005,JaehneNJP2009,prabl} and lasing~\cite{Lasing, Lasing2, quantum-dots}  have previously been discussed.
The resulting cooling rate is optimized by choosing a laser detuning $\delta_L=-\omega_m$ [see Fig.~\ref{fig:Setup} b)] and scales approximately as  
\begin{equation}\label{eq:Gamma_z}
\tilde{\Gamma}_\parallel \approx \frac{\lambda_\parallel^2}{\Gamma} \frac{ \Omega^2}{ \omega_m^2},
\end{equation}
where $\Omega$ is the Rabi frequency. 
For this off-resonant coupling, 
we see that the phonon sideband transitions are suppressed at the large mechanical frequencies
typical of diamond nanostructures. 
In contrast, for the second type of coupling, $\sim \Sigma_\perp(a+a^\dag)$, the mechanical frequency can be compensated by matching the frequency spacing $\Delta$ between the states $|x\rangle$ and $|y\rangle$, leading to resonant cooling and heating process 
indicated in Fig.~\ref{fig:Setup} c) and d). 
The corresponding rates are optimized for $\Delta=\pm \omega_m$ and with
resonant laser driving, $\delta_L=0$. The resulting scaling is 
\begin{equation}\label{maxLDx}
\tilde{\Gamma}_\perp \approx \frac{\lambda_\perp^2}{\Gamma} \frac{ 4 \Omega^2 }{\Gamma^2},
\end{equation}
and shows that the cooling and heating rates can be maximized independent of the mechanical frequency, by saturating the excited states, $\Omega\approx \Gamma$. This difference in the scaling has important practical implication when the laser power is limited by heating of the sample or by two-photon charging effects~\cite{BehaPRL2012,SiyushevPRL2013}. Therefore,  the near degenerate excited state manifold of the NV defect could provide a crucial ingredient for a first experimental demonstration of strain induced cooling and lasing effects for nanomechanical systems.

\subsection{Thermometry and probing NV-phonon interactions}

The same mechanisms outlined above for manipulating the state of a 
mechanical resonator can also be used for readout. 
For example, as discussed in  detail in Sec.~\ref{sec:Spectrum}, by exciting the state $|y\rangle$ with a $y$-polarized laser, the total photon flux $I_x$ of the $x$-polarized light scattered from state $|x\rangle$ is directly proportional to the phonon occupation number,
\begin{equation}\label{eq:Ix}
I_x(\delta_L)\approx  \frac{4 \lambda_\perp^2  \Omega^2 \Gamma}{(\Gamma^2 + 4\delta_L^2)^2} \times \langle a^\dag a \rangle.
\end{equation}
This provides an efficient and, ideally, noise-background-free way to directly measure the effective temperature of the mechanical mode.

Finally, we emphasize that the  manipulation and readout schemes 
discussed in this work for a single mechanical mode can serve as a versatile set of tools for investigating the still poorly understood nature of NV-phonon interactions. For example, in Sec.~\ref{sec:multiphonon} we identify multi-phonon lasing effects which result from the interplay between both $\Sigma_\parallel$ and $\Sigma_\perp$ type interactions. 
For studies such as this, the strong coupling to a resonant mode and the ability to amplify the mode using phonon lasing  could provide much cleaner experimental signatures than looking at similar effects in bulk diamond~\cite{PRL2009}.

\section{Model}\label{sec:Model}
We now proceed with a more detailed derivation of the strain coupling of an NV center to a single vibrational mode, 
taking into account the multi-level structure mentioned above.
Similar models of phonon coupling to a single excited state~\cite{HabrakenNJP2012,Albrecht2013} and direct phonon coupling 
to the spin sub levels of the NV electronic ground state manifold~\cite{spin-spin} have been discussed
previously.

\subsection{Level structure and strain coupling of a NV center in diamond}

\begin{figure}
 \begin{center}
\includegraphics[width=0.48\textwidth]{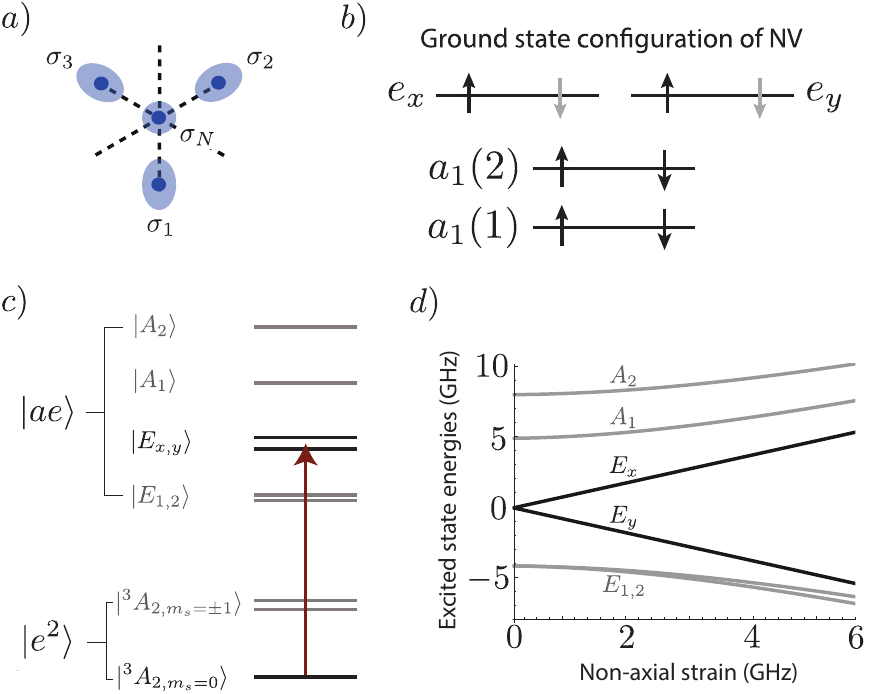}
  \caption{(Color online). a) Schematic top view of the NV defect and its dangling bond representation. The shaded areas depict the (hybridised) sp$^3$ bonding orbitals ($\sigma_{1,2,3}$ for the carbon and $\sigma_N$ for the nitrogen atoms). b) Ground state single particle configuration in the electron (black) and in the hole (empty arrows) representations. c) Energy level diagram of the NV center showing the spin sub-levels of the ground and the first excited triplet state. The relevant three-level structure used in this work is highlighted in black. d) Excited state energy splitting induced by the non-axial strain~\cite{NatureLukin,Group,Group2}.}
  \label{fig:ElectronicLevels}
 \end{center}
 \end{figure}

 The negatively charged NV$^-$ color center in diamond is formed by a substitutional nitrogen atom and an adjacent lattice vacancy. As shown in Fig.~\ref{fig:ElectronicLevels} a), the center has a
$C_{3v} $ symmetry and the six outer electrons occupy four orbitals 
labeled $a_1(1),a_1(2),e_x,e_y $. 
These are linear combinations of the ``dangling bond'' electronic orbitals located at the carbon and the nitrogen atoms, and they transform as the irreducible representations of the symmetry group~\cite{Group,Group2}. In the ground state, four electrons occupy the fully symmetric orbitals $a_1(1), a_1(2)$. The remaining two occupy the degenerate $e_x,e_y$ orbitals, forming a spin triplet which minimizes the electron-electron Coulomb interactions~\cite{Group}. Equivalently, this state can be described in terms of two holes occupying the levels $e_x$ and $e_y$, indicated by
the empty arrows in the Fig.~\ref{fig:ElectronicLevels} b).
Adopting this hole representation, the ground state is conventionally denoted by~\cite{Group} 
%
%
\begin{equation}\label{eq:GS}
|^3A_{2 m_s}\rangle = | e_xe_y-e_ye_x\rangle \otimes | m_s\rangle,
\end{equation}
 %
%
%
where $ m_s =\pm1,0 $ labels the three possible spin projections
and the degeneracy of states with $|m_s|=1$ and $m_s=0$
is lifted by a zero field splitting of 
 $\sim$ 2.87 GHz due to  spin-spin interactions.  
The first electronically excited state is $1.95$ eV higher in energy and corresponds to the promotion of one hole to the $a_1(2)$ orbital. 
This orbital doublet combined with the spin triplet 
yields six states in the excited state manifold
(the $ae$ configuration),
labeled by
$| E_{1,2} \rangle$,$| E_{x,y} \rangle$ and $| A_{1,2} \rangle$. 
These states are separated in energy by a few GHz due to spin-orbit and spin-spin interactions~\cite{Group,Group2},
and the resulting level ordering is shown in Fig.~\ref{fig:ElectronicLevels} c).
In this work we are mainly interested in the two excited states 
with zero spin angular momentum
\begin{equation}\label{eq:exc_states}
| E_{x,y} \rangle = | ae_{x,y} - e_{x,y} a\rangle \otimes | m_s=0 \rangle,
\end{equation}
which can be selectively excited by linearly polarized light from the $m_s=0$ ground state. Due to their vanishing spin projection number, $|E_x\rangle$ and $|E_y\rangle$ are not mixed with the other levels by spin-orbit interactions and they are degenerate in the absence of strain or external electric fields. 
For simplicity, from here on we will adopt the shorthand notation $|x\rangle\equiv |E_x\rangle$, $|y\rangle\equiv |E_y\rangle$ for the two excited states and  $|g\rangle \equiv |^3A_{2 0}\rangle$ for the ground state.

\subsection{Strain and NV-phonon interactions}

The effect of strain on the electronic states can be described by a deformation potential coupling $H_{\rm strain} = H_{\rm a} +H_{\rm na}$. Here, the axial part $H_{\rm a}$ accounts for those lattice deformations which are totally symmetric (of $A$-type as those belonging to the $A_1$ irreducible representation of the point group), while the non-axial part $H_{\rm na}$ arises from deformations which break the $C_{3v}$ symmetry ($E$-type)~\cite{Jahn-Teller,PRL2009,NV-Phonon}.
Since in the ground state  there is only a single electronic orbital, the state $|g\rangle$ is highly immune against lattice distortions and the effect of $H_{\rm strain}$ on $|g\rangle$ can be neglected (for a higher order effect of strain on the $m_s=\pm1$ spin levels see Ref.~\cite{spin-spin}). In contrast, the degeneracy between the $e_x$ and $e_y$ orbitals makes the excited states highly susceptible to external perturbations~\cite{Group,Group2}. Projected onto our states of interest, $|x\rangle$ and $|y\rangle$, the resulting strain coupling is 
\begin{equation}\label{eq:Ha}
H_{\rm a}=\epsilon_A\Xi_A   \left(|x\rangle \langle x| + |y\rangle\langle y|\right),
\end{equation}
for the axial part and 
 \begin{equation}\label{eq:Hna}
\begin{split}
H_{\rm na} = &\epsilon_{E} \Xi_{E} \left(|x\rangle\langle x| - |y\rangle\langle y|\right)+ \epsilon_{E}^\prime \Xi_{E}^\prime  \left(|x\rangle\langle y| + |y\rangle\langle x|\right),
\end{split} 
\end{equation} 
for the non-axial part, respectively. Here  $\Xi_A$, $\Xi_E$ and $\Xi_E^\prime$ are deformation potential constants and $\epsilon_A$, $\epsilon_E$ and $\epsilon_E^\prime$
denote the appropriate components of the strain tensor, which can be derived from group theoretical considerations \cite{Group}. While $H_{\rm a}$ preserves the symmetry of the electronic states and therefore only shifts the energy of the excited states relative to the ground state, 
the two contributions in $H_{\rm na}$ account for a strain induced splitting of $|x\rangle$ and $|y\rangle$ relative to each other as well as a strain induced mixing between the two excited states. The displacements, and phonons, that couple in this 
way are transverse to the NV axis, and correspond to E-type symmetry.



We are interested in the strain field associated with the quantized vibrational modes of the nanobeam.
For small displacements the induced strain at the position of the NV center is linear in the mode amplitudes and in second quantization the strain Hamiltonian given in Eqs.~\eqref{eq:Ha} and ~\eqref{eq:Hna} can be written in the form~\cite{PRL2009}
\begin{equation}\label{eq:QStrain}
H_{\rm strain} =  \hbar \left(\sum_n  \sum_{\nu=0,\parallel,\perp} \lambda^n_{\nu} \Sigma_{\nu} \right) (a_n+a_n^{\dagger}).
\end{equation}
Here $a_n$  and $a_n^\dag$ are the bosonic operators for the $n$-th vibrational mode, $\lambda^n_\nu$ are the corresponding coupling constants.  The operators $\Sigma_\parallel$ and $\Sigma_\perp$ have been defined below Eq.~\eqref{eq:HNVphon} and here we have also included $\Sigma_0 = |x\rangle\langle x| + |y\rangle\langle y|$ to account for a common shift of the excited states due to axial strain. 
%
In micron-sized diamond structures the mode frequencies $\omega_m$ are separated by a few GHz, which in our analysis below allows us to restrict Eq.~\eqref {eq:QStrain} to a single near-resonant mode with a mechanical vibration frequency $\omega_m$ and bosonic operator $a$. 
The values of the corresponding coupling parameters $\lambda_0$, $\lambda_\parallel$ and $\lambda_\perp$ depend on details of the specific experimental setup, such as the resonator dimensions, the vibrational mode function of interest as well as the orientation of the NV center in the diamond lattice. In the following it is assumed that there is no `accidental' symmetry and that all the $\lambda_\nu$ are  similar in magnitude.
 
To estimate the absolute strength of the NV-phonon coupling we consider a doubly clamped diamond nanobeam of dimensions $(l,w,t)=(2,0.2,0.2) \,\mu$m. The fundamental bending mode of this beam has a frequency of $\omega_m/(2\pi)\approx 1$ GHz. For a NV center positioned at distance $z_0$ away from the axis of the beam, the induced stress per zero point oscillation $a_0$ is approximately given by $\sigma = [ \partial^2u(x)/\partial x^2] E z_0a_0$, where $E\approx 1.2$ TPa is the Young's modulus and $ u(x) $ is the displacement field of the fundamental mode~\cite{WilsonRaePRL2004,spin-spin,Cleland}. Measurements of the NV energy level splitting as a function of applied stress~\cite{vibr-diamond} give values around $\partial \omega/\partial \sigma\sim 2\pi\times 1$ kHz. This  corresponds to a deformation potential coupling of $\Xi\approx 5$ eV and $\lambda/(2\pi) \approx 6$ MHz. Similarly, by considering the lowest order compression mode (along the long axis of the beam) we obtain a mechanical frequency of $\omega_m/(2\pi)\approx 4.5$ GHz. In this case the stress per zero-point motion is given by $ \sigma =  [\partial u(x)/\partial x] Ea_0$, where $u(x)= \sin(\pi x/L)$, and results in a similar coupling constant of $\lambda/(2\pi)\approx 6.5$ MHz. These estimates show that in micron scale structures NV-phonon couplings of a few MHz are expected, while, for example, by using a compression mode, the NV center is still located sufficiently far from the surface. 


\subsection{Laser driving and dissipation}
For the cooling and lasing effects discussed below we assume that the NV center is driven by a near resonant laser of frequency $\omega_L$. For concreteness we assume that the excitation laser is linearly polarized along the $y$ axis and detuned from the state $|y\rangle$ by $ \delta_L$. In the frame rotating with the laser frequency the resulting effective model Hamiltonian for our system is $(\hbar=1)$
%
\begin{equation}\label{eq:Hmodel}
\begin{split}
	H= &\omega_m a^\dag a  - \delta_L |y\rangle\langle y| - 
	(\delta_L - \Delta) |x\rangle\langle x|  \\
	&+  \frac{\Omega}{2} (|y	
	\rangle \langle g| + |g \rangle \langle y |) + \lambda \bar{\Sigma} (a+a^{\dagger}),
\end{split}
\end{equation}
where we have introduced the short notation $\lambda \bar \Sigma \equiv \sum_{\nu=0,\parallel,\perp} \lambda_{\nu} \Sigma_{\nu}$ 
and $\Delta=\omega_x-\omega_y \sim 1$ GHz is the frequency splitting between the two excited states $|x\rangle$ and $|y\rangle$ due to static lattice distortions. This splitting  can be tuned by applying external electric fields~\cite{TamaratPRL2006}
and in the following we treat $\Delta$ as an adjustable parameter. 

To account for dissipation due to radiative and mechanical losses we model the system dynamics by the master equation 
\begin{equation}\label{ME}
\dot{\rho} = -i \left[ H,\rho \right] + \mathcal{L}_{\Gamma}\rho  + \mathcal{L}_{\gamma}\rho,
\end{equation}
for the system density operator $\rho$. The Liouville operator $\mathcal{L}_{\Gamma}$ is given by
\begin{equation}\label{relax} \begin{split}
\mathcal{L}_{\Gamma}\rho = & \frac{\Gamma}{2} \sum_{\xi=x,y}( 2 |g\rangle\langle \xi|\rho|\xi\rangle\langle g| - |\xi\rangle\langle\xi| \rho - \rho|\xi\rangle\langle\xi|)\\
& + \frac{\Gamma_\phi}{2} \sum_{\xi = x,y} ( 2 |\xi\rangle\langle \xi|\rho|\xi\rangle\langle \xi| - |\xi\rangle\langle\xi| \rho - \rho|\xi\rangle\langle\xi| ),
\end{split} \end{equation}
and describes the radiative decay of the excited states with an approximately equal decay rate $\Gamma/(2\pi)\approx 15$ MHz as well as an additional broadening $\sim \Gamma_\phi$ of the optical transitions  due to spectral diffusion. 
In bulk diamond and low temperatures of $T<10$ K, 
narrow optical lines with $\Gamma_\phi \sim \Gamma$
can be achieved~\cite{nol1,nol2}.
For shallow implanted NVs, surface impurities induce additional dephasing and 
significant experimental effort is devoted to
understanding and mitigating this additional dephasing. 
For NV centers located a few tens of nanometers away from the surface, 
it is expected that 
sufficiently narrow lines with $\Gamma_\phi \lesssim 100$ MHz can be reached.

The last term in Eq.~\eqref{ME} describes 
mechanical dissipation due to the coupling 
of the resonant vibrational mode to the thermal bath of phonon modes in the support.  
It is given by 
\begin{equation}
\begin{split}
\mathcal{L}_{\gamma}\rho = &  \frac{\gamma}{2} ( N_{th}+1 ) \mathcal{D}[a]\rho + \frac{\gamma}{2} N_{th} \mathcal{D}[a^\dag]\rho,
\end{split}\end{equation}
where $\mathcal{D}[a]\rho=( 2a\rho a^{\dagger} - a^{\dagger}a\rho - \rho a^{\dagger}a )$,  $\gamma = \omega_m / Q $ is the mechanical damping rate for a vibrational mode of quality factor $Q$ and $ N_{th} = (e^{\hbar \omega_m / k_B T} - 1 ) ^{-1} $ is the equilibrium phonon occupation number for a  support temperature $T$. For mechanical frequencies $\omega_m/(2\pi)\approx 1$ GHz and realistic values of $Q\approx 10^5-10^6$~\cite{Ovartchaiyapong2012,Tao2012} the corresponding to damping rates are a few kHz and $N_{th}\approx 100$ at $T=4$ K.

\section{Cooling}\label{sec:Cooling}
In Sec.~\ref{sec:Idea} we have outlined the basic idea, how in the present system phonon-assisted processes depicted in Fig.~\ref{fig:Setup} b)-d) can lead to cooling and heating. In the following we will first focus on the cooling effects induced by the $\sim \Sigma_\parallel$ and $\sim \Sigma_\perp$ type interactions
and evaluate the conditions for ground state cooling of the mechanical mode.

\subsection{Effective cooling equation}\label{cool-eq}
For the parameters of interest $\lambda < \Gamma$ and low mechanical occupation numbers, the dynamics of  the NV center is only weakly perturbed by the phonon mode. This allows us to adiabatically eliminate the NV center degrees of freedom and derive an effective equation of motion for the mechanical degrees of freedom only~\cite{Cirac,WilsonRaePRL2004,JaehneNJP2009,ZippilliPRL2009,prabl}. To do so, we change into a frame rotating with $\omega_m $ and decompose the ME~\eqref{ME} into three terms,
\begin{equation}
\dot{\rho} =  \mathcal{L}_{\rm NV}\rho + \mathcal{L}_{\lambda}\rho + \mathcal{L}_{\gamma}\rho,
\end{equation}
where  $\mathcal{L}_{\rm NV}$ and $\mathcal{L}_\gamma$ describe the bare dynamics of the NV center and the intrinsic dissipation of the mechanical mode, respectively. Finally, $\mathcal{L}_\lambda$ accounts for the coupling between the NV center and the mechanical mode, which in the rotating frame is given by
\begin{equation} 
\mathcal{L}_{\lambda}\rho = - i \lambda [\bar \Sigma (a e^{-i\omega_mt}Ê+a^\dag e^{i\omega_m t}),\rho].
\end{equation}
In the limit $\lambda\rightarrow 0$, the defect and the phonon mode are decoupled and the system relaxes into the state $\rho(t)\simeq \rho_0\otimes \rho_m(t)$, where $\rho_0$ is the steady state of the driven NV center defined by $\mathcal{L}_{\rm NV}\rho_0=0$ and $\rho_m(t)$ is the reduced density operator of the mechanical mode. 
Provided the condition $ \gamma N_{th},Ê\lambda \sqrt{\langle n\rangle + 1/2} \ll \Gamma, \omega_m$ is satisfied, 
where $\langle n\rangle$  is
the mean occupation number of the mechanical mode,
the effect of $\mathcal{L}_\lambda$ can be treated in perturbation theory. 
%
%
Using a projection operator method we derive an effective master equation for the mechanical mode~\cite{Cirac,prabl,JaehneNJP2009}
\begin{equation}\label{eq:CoolingME}
\begin{split}
\dot{\rho}_m = & \mathcal{L}_{\gamma} \rho_m + \frac{\tilde{\Gamma}}{2}(N_0+1)\mathcal{D}[a]\rho  + \frac{\tilde{\Gamma}}{2}N_0     \mathcal{D}[a^\dag ]\rho.
\end{split}\end{equation}
Here we have introduced the cooling rate $ \tilde{\Gamma} = 2 \lambda^{2} ({\rm Re}[S(\omega_m)] - {\rm Re}[S( -\omega_m)]) $ and the minimal occupation number $ N_0 = {\rm Re}[S(-\omega_m)]/({\rm Re}[S(\omega_m)] - {\rm Re}[S(-\omega_m)])$, which are determined by the equilibrium fluctuation spectrum
\begin{equation}\label{spectrum}
S(\omega_m) = \int_0^{\infty} dt^{\prime} \langle \bar {\Sigma}(t^{\prime}) \bar {\Sigma} (0)\rangle e^{i\omega_m t^{\prime}},
\end{equation}
where $\langle \cdot\rangle$ denotes the average with respect to the stationary NV center state $\rho_0$.
This spectrum can be evaluated using the quantum regression theorem~\cite{Walls, Lambro} and the main steps of this calculation and the general result are  summarized in App.~\ref{Spec}.

From Eq.~\eqref{eq:CoolingME},
the mean occupation number $\langle n\rangle=\langle a^\dag a\rangle$ of the phonon mode
satisfies 
\begin{equation}
\partial_{t} \langle n \rangle = - \tilde \Gamma \left( \langle n \rangle - n_{f} \right),
\end{equation}
where for $\tilde \Gamma \gg \gamma$ and $N_{th}\gg 1$ the final occupation number $n_f$ is approximately given by
 \begin{equation}
 n_{f}\approx \frac{\gamma N_{th}}{\tilde{\Gamma}}+N_0.
\end{equation}
In the following discussion we are mainly interested in the sideband resolved regime $\Gamma,\Gamma_\phi\ll \omega_m$ where $N_0 \ll 1$ can be neglected. The final mode occupation number is then determined by the competition between the optical cooling rate $\tilde \Gamma$ and the rethermalization rate $\gamma N_{th}$.

\subsection{Results and discussion}

\begin{figure}
 \begin{center}
\includegraphics[width=0.48\textwidth]{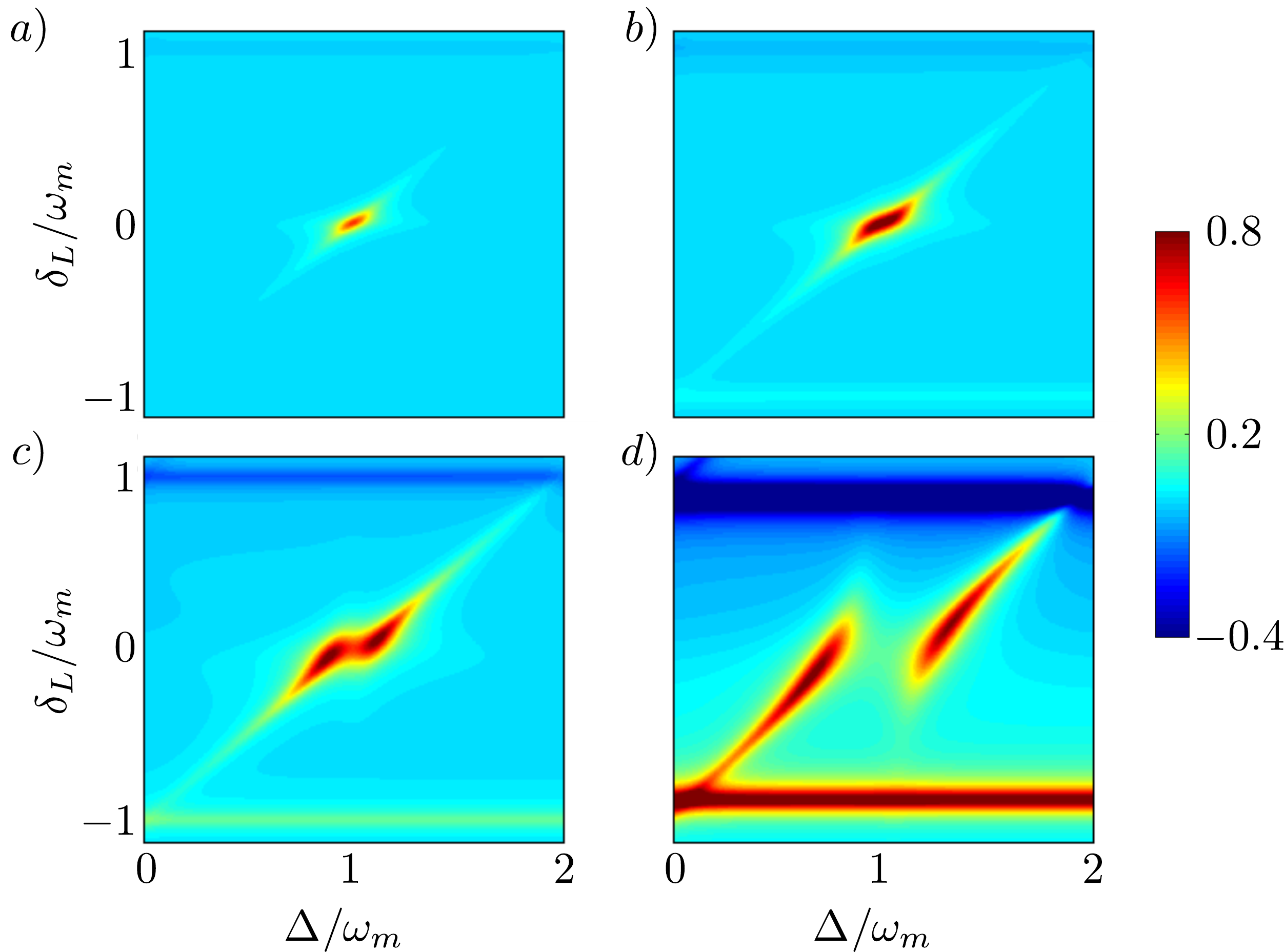}
  \caption{  (Color online). Density plots of the Lamb-Dicke cooling rate (in units of $ \lambda^2/\Gamma $) as a function of the detuning $\delta_L$ (y-axis) and the frequency difference $\Delta$ of the excited levels (x-axis) for four different values of the Rabi frequency: a) $ \Omega/\Gamma = 0.5 $, b) $  \Omega/\Gamma = 1 $, c) $  \Omega/\Gamma = 2 $ and d) $ \Omega/\Gamma = 5 $. For all plots it has been assumed that $ \lambda_{\perp}=\lambda_{\parallel}=\lambda $, $\lambda_0=0$ and $\Gamma_\phi=0$.}
 \label{fig:CoolingRates}
 \end{center}
 \end{figure}
 
In Fig. \ref{fig:CoolingRates} we numerically evaluate the cooling rate $ \tilde{\Gamma} $ and plot the result as a function of $ \Delta $ and $ \delta_L $ and different values of the driving strength $\Omega$. We find regions of strong cooling around $\delta_L\approx -\omega_m$ and around $\delta_L\approx0$, $\Delta\approx\omega_m$, which can be associated with the two excitation processes indicated in Fig. \ref{fig:Setup} b) and c), respectively. In the first case the laser is tuned on the red sideband of the $|g\rangle\rightarrow|y\rangle$ transition and a mechanical energy of $\hbar \omega_m$ is absorbed to make this transition resonant. In the second case the laser excites the state $|y\rangle$ on resonance, and by absorbing an additional phonon, the NV center is further excited to the state $|x\rangle$ before it decays. For large $\Omega>\Gamma$, the cooling maximum is separated into two peaks as a result of the strong Rabi splitting. 

Fig.~ \ref{fig:CoolingRates} shows that while at larger driving powers $\Omega\approx \omega_m$ both cooling mechanisms lead to appreciable rates of $\tilde \Gamma \sim \mathcal{O}(\lambda^2/\Gamma)$, the mechanism related to $\Sigma_0$- or $\Sigma_{\parallel}$-type coupling is strongly reduced at lower Rabi frequencies. To see this more explicitly we evaluate the cooling rate $\tilde \Gamma$ under weak-driving conditions ($ \Omega < \Gamma ,\omega_m$) and for the two types of couplings $\sim \Sigma_{\parallel} $ and $\sim \Sigma_{\perp} $ separately. 
In the first case we obtain
\begin{equation}\begin{split}
\tilde{\Gamma}_{\parallel} = \frac{ 4 \Gamma \lambda_{\parallel}^2  \Omega^2 }{\Gamma^2 + 4\delta_L^2} & \left[ \frac{1}{\Gamma^2 + 4(\omega_m+\delta_L)^2} \right. \\
& - \left. \frac{1}{\Gamma^2 + 4(\omega_m -\delta_L)^2} \right],
\end{split}\end{equation}
in agreement with previous results for phonon cooling schemes with two level systems~\cite{prabl}. For sideband resolved conditions, $ \Gamma \ll \omega_m $,  this cooling rate is optimized for $ \delta_L = -\omega_m $ and with a maximal value given by Eq.~\eqref{eq:Gamma_z} in Sec.~\ref{sec:cooling-lasing}. On the other hand, by considering only the $ \Sigma_x $ coupling we obtain
\begin{equation}\begin{split}\begin{split}
 \tilde{\Gamma}_{\perp}  = \frac{ 4 \Gamma \lambda_{\perp}^2 \Omega^2}{\Gamma^2+4\delta_L^2} &  \left[ \frac{1}{\Gamma^2 + 4(\Delta - \omega_m -\delta_L)^2} \right. \\
 & - \left. \frac{1}{\Gamma^2 + 4(\Delta + \omega_m -\delta_L)^2} \right].
\end{split}\end{split}\end{equation}
Again under side-band resolved conditions, the maximal rate in this case occurs for $ \delta_L = 0 $ and $ \Delta = \omega_m $, where the maximal value is given by Eq.~\eqref{maxLDx} in Sec.~\ref{sec:cooling-lasing}.
We see that the requirement to maximize the cooling rate
is now only
$\Omega\sim \Gamma$, which corresponds to a saturation of the state $|y\rangle$ on resonance. This is a significant  improvement compared to the
much stronger requirement $\Omega \sim \omega_m$ in Eq.~\eqref{eq:Gamma_z}
when the mechanical frequency is high, $\omega_m \gg \Gamma$.
For example, by comparing Eqs.~\eqref{eq:Gamma_z} and \eqref{maxLDx} for typical parameters considered in this work and assuming  $\lambda_{\perp} \sim \lambda_{\parallel}$, we find that the optimal cooling rate 
for the same $\Omega$
is improved by a factor
\begin{equation}
\frac{ \tilde{\Gamma}_{\perp}}{ \tilde{\Gamma}_{\parallel} } \approx  \frac{4\omega_m^2}{\Gamma^2} \approx 10^4.
\end{equation}
%
In other words, the laser power that is needed to achieve the same cooling rate can be a factor $10^4$ lower when making use of the multi-level structure of the NV center. This is an important practical issue at low temperature where absorbed laser light might otherwise lead to heating of the entire sample.

\subsection{Ground state cooling} 

As mentioned above, the final occupation number $n_f$ in  the sideband resolved regime is mainly determined by the competition between the cooling rate $\tilde \Gamma$ and the rethermalization rate $\gamma N_{th}\simeq k_BT/(\hbar Q)$. Under optimal driving the maximal achievable cooling rate approaches $ \tilde \Gamma^{\rm max}\approx \bar \lambda^2/\Gamma $. This happens for laser powers $ \Omega \sim \omega_m $ for the $ \Sigma_{\parallel} $-type coupling and for $ \Omega \sim \Gamma $ for the $ \Sigma_{\perp}$-type coupling. The minimal achievable occupation numbers are then approximately given by  $ n_f\approx \gamma N_{th} \Gamma/\lambda^2$. For $\lambda/(2\pi)\approx 5$ MHz, ground state cooling $n_f\lesssim 1$ can be achieved for realistic mechanical quality factors of $Q\approx 10^5$  and initial temperature of $T=4$ K.

In our analysis so far we have considered the ideal case of purely radiatively broadened optical lines $\Gamma > \Gamma_\phi$, which is a realistic assumption in bulk diamond and at temperatures of a few Kelvin. In nanoscale structures, noise processes on the surface become important and can lead to additional spectral diffusion of the optical line. For the cooling to remain efficient, we require that $\Gamma_\phi< \omega_m$, such that the phonon sidebands are still well resolved.  
Based on rapid progress 
with shallow-implanted NVs and expected line widths of $\Gamma_\phi\sim 200$ MHz,
this condition can be realistically achieved for $\sim$ GHz mechanical modes.  
Since spectral diffusion broadens the line without causing dissipation, the 
cooling rate is reduced by a factor $\tilde \Gamma \sim \Gamma/(\Gamma_\phi+\Gamma)$. 
This slightly degrades the cooling, but does not affect the mechanism itself.

It is important to point out that in our model in Eq.~\eqref{relax} a simple Markovian linebroading $\sim \Gamma_\phi$ is assumed. In practice the spectral diffusion of the excited states is often better described by a highly non-Marokvian, slow drift of the excited state energies. This can in principle be compensated by applying additional optical dressing or real-time feedback schemes to stabilize the optical transitions and a reduction of the remaining broadening to $\Gamma_\phi\sim \Gamma$ seems feasible.


\section{Phonon Lasing}\label{sec:Lasing}
As a second application we now consider the opposite regime, where the detuning of the optical driving field is chosen to enhance phonon emission processes. At low driving powers this simply leads to an increase of the mechanical energy, but at larger driving strengths the heating can overcome the intrinsic mechanical damping and drive the resonator into a large amplitude coherent state. In analogy to a strongly pumped optical mode undergoing a lasing transition, this effect is commonly referred to as `phonon lasing' and has been investigated in different physical settings~\cite{Bennett, Mficromaser,Lasing2,a-phonon-laser,phonon-laser-action,Lasing,quantum-dots,phonon-lasing}. While mechanical systems can in principle be driven into a coherent state by applying a resonant external force, this becomes increasingly more difficult for high frequencies modes in small structures.  
In contrast to the cooling mechanism discussed above, the phonon-lasing scheme
we now discuss amplifies the mechanical motion, providing an efficient way to probe NV-phonon interactions.

\subsection{Semiclassical phonon lasing theory}

In the previous section we derived an effective rate equation for the resonator mode under the assumption $ \lambda\sqrt{\langle n\rangle } \ll \Gamma $. In the opposite regime of amplification, the mean resonator occupation $\langle n\rangle$ can become very high and non-linear saturation effects -- which eventually limit the maximal achievable occupation number -- become important. Still assuming $\lambda\ll \Gamma$ these effects can be described within a semiclassical approach\cite{QuantumNoise}, where the effect of a large classical phonon amplitude $\sim \lambda \sqrt{\langle n\rangle}$ on the NV center dynamics is taken fully into account.  

Here we closely follow the phase-space approach, which was used in Ref.~\cite{prabl} to model phonon cooling effects at high initial temperatures. 
We introduce a set of quasi-probability distributions 
\begin{equation}\label{distributions}
P_{jk}(\alpha,t) = \frac{1}{\pi^2} \int  d^2\beta \, e^{\alpha \beta^* - \alpha^* \beta} {\rm Tr} \left\{e^{\beta a^{\dagger}} e^{-\beta^* a} \sigma_{jk} \rho(t)  \right\},
\end{equation}
where $ \sigma_{jk}= |j\rangle\langle k|$ and $j,k=g,x,y$. The $P_{jk}(\alpha,t)$ correspond to the expectation value of the operator $\sigma_{jk}$ for a fixed coherent state amplitude $\alpha$ and  $\langle \sigma_{jk} \rangle(t) =  \int P_{jk}(\alpha,t) d^2\alpha$.  The function $P(\alpha,t)=P_{gg}(\alpha,t)+P_{xx}(\alpha,t)+P_{yy}(\alpha,t)$ is the usual Glauber-Sudarshan P representation~\cite{Walls, Lambro,QuantumNoise} of the mechanical resonator density matrix.

In the frame rotating with $\omega_m$, the state of the mechanical mode changes slowly on the relaxation timescale $\Gamma^{-1}$ of the NV excited states. This allows us to evaluate the quasi-stationary values of $P_{ij}(\alpha,t)$ for a fixed point $\alpha$ in phase space, and insert the result back into the equation of motion for the P-representation $P(\alpha,t)$. In App.~\ref{FokkerPlanck} we use a Floquet expansion to apply this idea for the present system and derive and effective Fokker-Planck equation for the mechanical mode, 
\begin{equation}\label{eq:FokkerPlanck}
\begin{split}
\dot{P}(\alpha,t) \simeq &\frac{1}{2} \left( \frac{\partial}{\partial \alpha} \alpha \gamma(\alpha)  +  \frac{\partial}{\partial \alpha^*} \alpha^* \gamma(\alpha)\right) P(\alpha,t) \\
& + \gamma N_{th} \frac{\partial^2}{\partial \alpha \partial \alpha^*} P(\alpha,t),
\end{split}\end{equation}
where $ \gamma(\alpha) = \tilde{\Gamma}(\alpha) + \gamma $. In the limit $\alpha\rightarrow 0$ the energy-dependent damping rate  $\tilde{\Gamma}(\alpha)\equiv \tilde \Gamma(|\alpha|)$ reduces to $\tilde{\Gamma}$ defined below Eq.~\eqref{eq:CoolingME}, and must be in general evaluated numerically as described in App.~\ref{FokkerPlanck}. Note that in Eq.~\eqref{eq:FokkerPlanck} we have  neglected the influence of the NV center on the diffusion term. This is justified in the current regime of interest,  $N_{th} \gg 1$, but must be taken into account when studying lasing effects at low thermal occupation numbers $N_{th}\sim 1$.\cite{Lasing,Lasing2,Bennett,Micromaser}  

Eq.~\eqref{eq:FokkerPlanck}  preserves the radial symmetry of the initial thermal state; thus, by writing $ \alpha = r e^{i\phi}$, we can rewrite it in terms of a 
Fokker-Planck equation for the radial distribution,
\begin{equation}\begin{split}
\dot{P}(r,t) = & \frac{1}{2} \left( \frac{\partial}{\partial r} r + 1 \right) \gamma(r) P(r,t) \\
& + \frac{\gamma N_{th}}{4} \left( \frac{\partial^2}{\partial r^2} + \frac{1}{r} \frac{\partial}{\partial r}\right) P(r,t).
\end{split}\end{equation}
The steady-state solution of the radial equation is $ P(r,\infty) = \mathcal{N} e^{-\phi(r)} $, where $ \mathcal{N} $ is a normalization constant such that $ 2\pi\int_0^\infty r P(r) dr = 1 $ and
\begin{equation}\label{eq:Phi}
\phi (r) = \frac{2}{\gamma N_{th}}  \int_0^r  r^{\prime} \gamma(r^{\prime}) dr.
\end{equation}
In the absence of driving, $ \gamma(r) = \gamma $ and we obtain the thermal distribution function $ P(r,\infty) = e^{-r^2 / N_{th}} / (\pi N_{th}) $. For the cooling schemes described in Sec.~\ref{sec:Cooling}, we obtain $ \tilde{\Gamma}(r\rightarrow 0)=\tilde \Gamma >0$, but $\tilde \Gamma(r)$ decreases at larger values of $r$, where saturation effects set in and limit the cooling effect~\cite{prabl}. 
In the following, we are mainly interested in detuning such that for low occupations, $\gamma(r\rightarrow 0) < 0$ and energy is pumped into the mechanical mode. Again, due to saturation, this heating decreases at large oscillation amplitudes, where eventually $\gamma(r\rightarrow\infty) =\gamma>0$.   
%

\subsection{From heating to lasing}
\begin{figure}
 \begin{center}
  \includegraphics[width=0.45\textwidth]{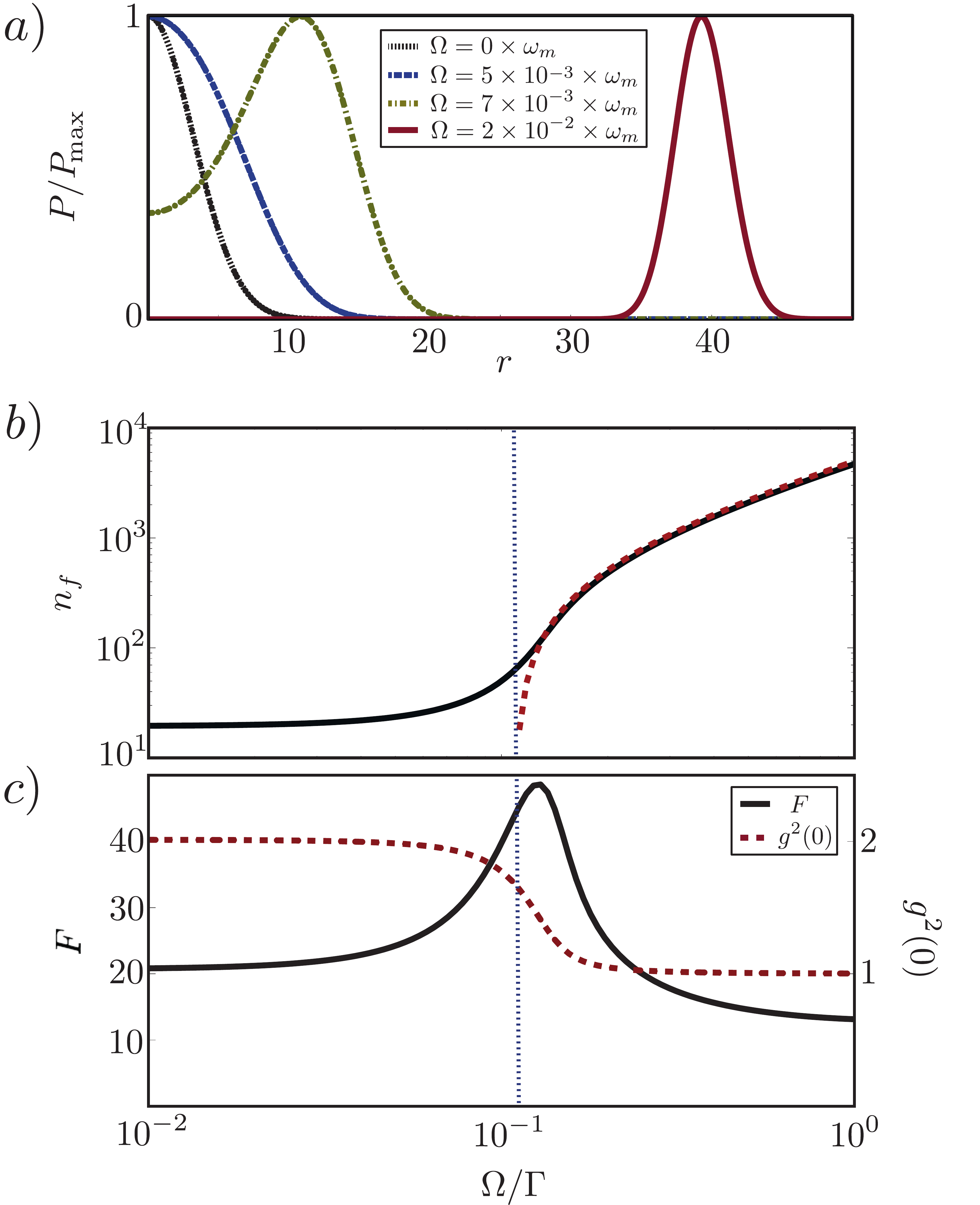}
  \caption{ (Color online). a) The stationary P-function $P(r,\infty)$ is plotted for different values of the Rabi frequency $\Omega$ given in the inset. Each curve is rescaled by its maximal value $P_{\rm max}$ and the other parameters used for this plot are (in units of $ \omega_m $), $ N_{th} = 20 $, $ \gamma = 10^{-6} $, $ \lambda_{\perp} = 0.001 $, $ \Gamma = 0.05 $ and $\Gamma_\phi=0$.  
 b) The final phonon occupation number $n_f$ is plotted as a function of $\Omega$ and other parameters as in a). The dashed line indicates the approximate result derived from the Gaussian P-function given in Eq.~\eqref{eq:PGauss}. c) Under the same conditions the Fano factor $F$ (solid line) and the correlation function $g^2(0)$ (dashed line) are plotted as a function of the driving strength. 
 In b) and c) the vertical dashed line indicates the position of the threshold given in Eq.~\eqref{eq:threshold}. 
 }\label{pfunctions}
 \end{center}
 \end{figure}
 
 In the previous section we have shown that resonant phonon interactions $\sim \Sigma_{\perp}$ provide an  efficient way to cool high frequency phonons, and in the following we analyze the reverse process of phonon lasing. To do so, we set $\Delta\approx-\omega_m$ and obtain the inverted  level structure shown in Fig. \ref{fig:Setup} d), where the driving laser excites the upper state $|y\rangle$, which can undergo a further transition to the lower state $|x\rangle$ by emitting a phonon.     
 
In Fig.~\ref{pfunctions} a) we present the numerically-calculated P-functions for different values of the driving strength $ \Omega$. For very low driving, the optical heating rate is still smaller than the intrinsic mechanical damping rate. In this case the resonator mode remains in a thermal state, but with a higher effective temperature. Above a threshold driving strength, $\Omega>\Omega_c$, the P-function starts to deviate from a thermal distribution and reaches its maximum at a finite value $r_0>0$. This is the onset of the lasing transition. By further increasing $\Omega$, the maximum shifts to larger and larger values and the P-function displays a narrow Gaussian shape, which approximates the sharp $\delta$-function, $P(r)\sim\delta(r-r_0)$, expected for an ideal coherent state. 
%
%

To further characterize the phonon lasing phenomenon, we plot in Fig.~\ref{pfunctions} b) the final phonon occupation number $ n_f $ as a function of $\Omega$, starting from an equilibrium value of $N_{th}=20$. We see that around $\Omega_c/\Gamma \approx 0.11$ the phonon number starts to increase significantly; for the chosen parameters, it can reach values up to $n_f \approx 10^4$. In Fig.~\ref{pfunctions} c) we show the corresponding values for  $g^2(0)=\langle a^\dag a^\dag a a\rangle/\langle n\rangle^2$ and the Fano factor $F=\langle n^2\rangle/\langle n\rangle$, which also show clear signatures of the transition from heating to lasing. For $\Omega  < \Omega_c$ the Fano factor remains close to $ F\approx n_f + 1 $, as expected for a thermal distribution. Above $\Omega_c$ the Fano factor starts to decrease, indicating a more Poisson-like distribution. This is even more apparent by looking at $g^2(0)$, which changes from a value of $g^2(0)=2$ for a thermal state to $g^2(0)\simeq1$ of a coherent state.

Note that an increase of the driving strength $\Omega> \Gamma$ leads to a saturation of the optical transition and therefore also the lasing effect. In addition, for a very strong driving field  $\Omega\gg\Gamma$, but otherwise fixed detunings, the resulting Rabi splitting between $|g\rangle$ and $|y\rangle$ will drive the system out of the resonance condition and the lasing effect breaks down.

Under weak driving conditions ($ \Omega < \Gamma, \omega_m $) and assuming a dominantly $ \Sigma_{\perp} $ coupling, 
we derive an approximate analytical form for the heating rate, which on resonance ($ \Delta = \omega_m $, $\delta_L=0$) is given by
\begin{equation}\label{eq:GxLasing}
\tilde{\Gamma}_{\perp}(r) = \frac{- 4 \lambda_{\perp}^2 \Gamma \Omega^2}{(\Gamma^2 + 4\lambda_{\perp}^2 r^2 )^2}.
\end{equation}
By direct integration of Eq.~\eqref{eq:Phi} we obtain
\begin{equation}
\phi (r) = \frac{r^2}{N_{th}} \left(  1 -  \frac{4 \lambda_{\perp}^2 \Omega^2}{\gamma \Gamma(\Gamma^2+4 \lambda_{\perp}^2 r^2) }  \right),
\end{equation}
and the position of the maximum of the P-function is found by solving $ \phi^{\prime}(r_0) = 0 $, 
\begin{equation}
r_0 = \frac{1}{2} \sqrt{ - \frac{\Gamma^2}{\lambda_{\perp}^2} + \frac{2\sqrt{\Gamma} \Omega}{\sqrt{\gamma} \lambda_{\perp}} }.
\end{equation}
Setting $ r_0 $ to zero yields the lasing-threshold,
\begin{equation}\label{eq:threshold}
 \frac{\Omega_c}{\Gamma} = \frac{\sqrt{\Gamma \gamma}}{2 \lambda_{\perp}},
\end{equation}
which is indicated in Fig.~\ref{pfunctions} by the vertical dotted line. 
Deep in the lasing regime, where $r_0\gg 1$, we can further make a saddle-point approximation 
and obtain a Gaussian P-distribution of the form
%
\begin{equation}\label{eq:PGauss}
P(r) \approx \frac{1}{ r_0 \sigma \sqrt{8\pi^3} } e^{-\frac{(r-r_0)^2}{2\sigma^2}},
\end{equation}
where the variance is given by $ \sigma^2 = 1/\phi^{\prime \prime}(r_0) $. From Eq.~\eqref{eq:GxLasing} we see that the requirement for lasing $|\tilde \Gamma_{\perp}(r\rightarrow 0)| \gg \gamma$ implies the condition $ \lambda_{\perp}^2\Omega^2 \gg \gamma \Gamma^3 $, for which the variance of the Gaussian distribution is essentially determined by thermal fluctuations, $ \sigma^2 \approx  N_{th} / 4 $. In this limit, the mean occupation number $ n_f \approx  r_0^2 + 3\sigma^2$ derived from Eq.~\eqref{eq:PGauss} is approximately given by
\begin{equation}
n_f \approx \frac{\Omega}{2\lambda_{\perp}} \sqrt{\frac{\Gamma}{\gamma}} + \frac{3}{4} N_{th}.
\end{equation}
Our analytical results are compared to the numerically-computed final phonon occupation number in Fig.~\ref{pfunctions} b), and we find very good agreement  above threshold.



\subsection{Cooling and lasing in the single- and multi-phonon regime}\label{sec:multiphonon}

\begin{figure}
 \begin{center}
\includegraphics[width=0.45\textwidth]{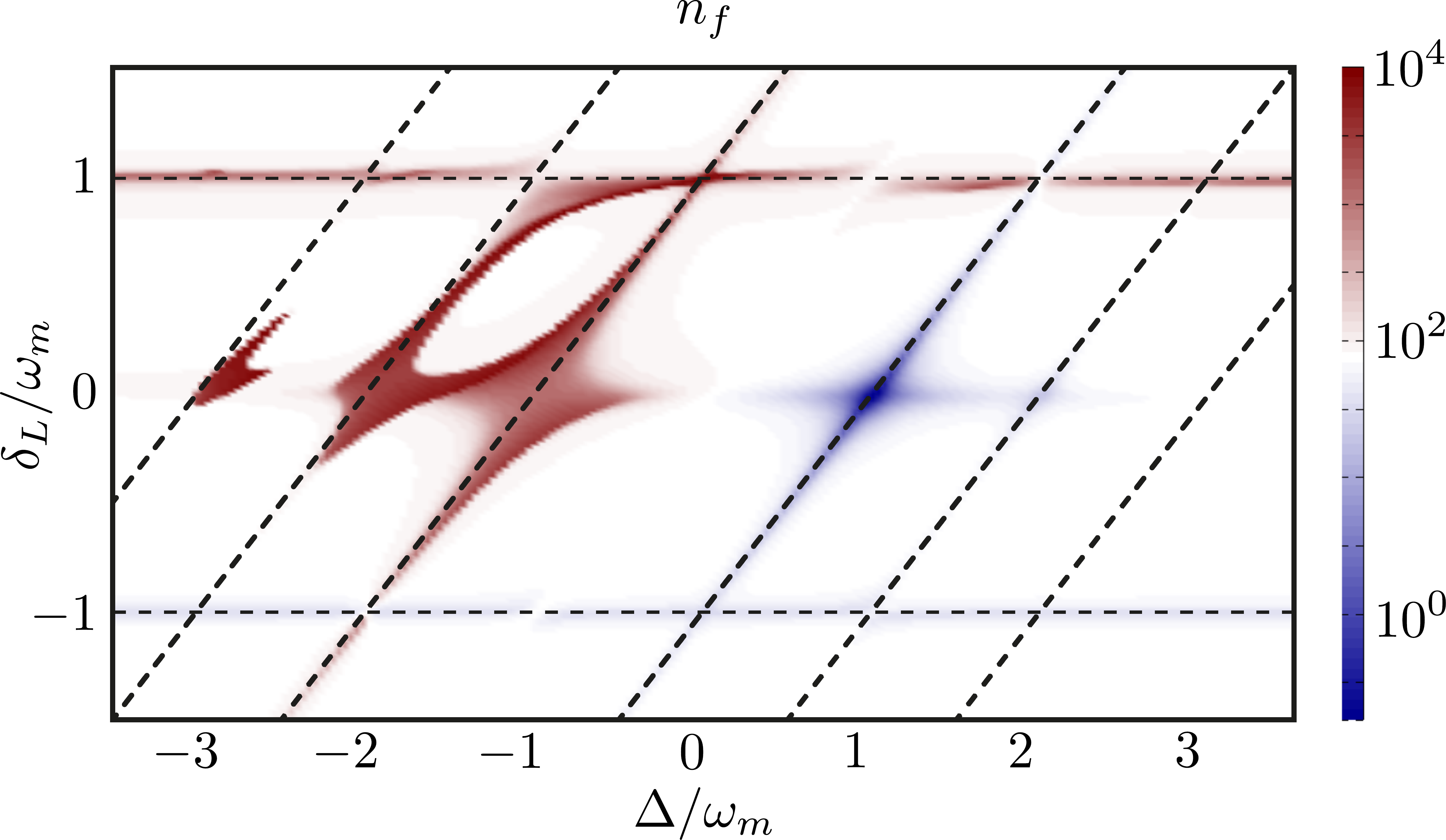}
  \caption{ (Color online). Numerically evaluated final phonon occupation number $n_f$ as function of $ \Delta $ and $ \delta_L $ and assuming an initial occupation of $ N_{th} = 80 $. The other parameters used for this plot are (in units of $ \omega_m $) $ \Omega = 0.05 $, $ \Gamma = 0.05 $, $ \gamma= 10^{-6} $, $ \lambda_{\perp} = \lambda_{\parallel} = 0.005 $ and $\Gamma_\phi=0$. The dashed lines indicate the resonance conditions for single and multi-phonon sidebands.  }\label{multi-phonon}
 \end{center}
 \end{figure}

In general, the presence of both $\Sigma_{\perp}$- and $\Sigma_{\parallel}$-type NV-phonon interactions can lead to a rich interplay between cooling and heating mechanisms, as
different single and multi-phonon processes become resonant
depending on the laser detuning $\delta_L$ and the excited state splitting $\Delta$.
This is illustrated in Fig.~\ref{multi-phonon}, where we evaluate numerically the final phonon occupation number $ n_f $ for a large range of detunings $\delta_L$ and $\Delta$. The plot shows  the same cooling and heating processes discussed above, corresponding to $\Sigma_{\perp}$-type (maximized for $\Delta=\pm\omega_m$, $\delta_L=0$) and $\Sigma_{\parallel}$-type (maximized for $\delta_L=\pm\omega_m$) interactions and associated with emission or absorption of single phonons.
In addition, we observe heating and cooling features at multiple integers of the phonon frequency, i.e. under the condition $ \delta_L - \Delta = \pm n \omega_m $, indicating multi-phonon processes. These effects are most pronounced in the lasing regime, where the mechanical mode is highly excited and higher order phonon-processes become relevant.  Note that such multi-phonon effects (for example the two- and three-phonon lasing peaks at $\Delta=-2\omega_m$ and $\Delta=-3\omega_m$) appear only in the presence of \emph{both} types of couplings. 
Similarly, two types of NV-phonon interactions are thought to be involved in
the NV zero-phonon line broadening and its 
$T^5$ scaling \cite{PRL2009}.
In light of this, studying multi-phonon lasing may provide a useful tool to analyze the detailed nature of NV-phonon coupling.


\section{Excitation Spectrum}\label{sec:Spectrum}
In this last section we study the excitation spectrum of the NV center, which provides a direct way to probe the state of the mechanical resonator by measuring the light scattered from the NV center. By considering a polarization-selective photon detection setup, we calculate the photon flux $I_{\eta=x,y}(\delta_L)=\Gamma \langle \sigma_{\eta\eta}\rangle$ emitted from the two excited states and as a function of the laser detuning $\delta_L$. According to the definition in Eq.~\eqref{distributions} we obtain
\begin{equation}
I_{\eta}(\delta_L)  =\Gamma  \int d^2 \alpha   \, P_{\eta\eta} (\alpha), 
\end{equation}
and under the validity of our semiclassical approximation, $P_{\eta\eta}(\alpha)\simeq X^0_\eta(\alpha) P(\alpha,\infty)$. Here $X^0_\eta(\alpha)$ is an energy-dependent factor, defined in Eq.~\eqref{eq:Xn} in App.~\ref{FokkerPlanck}, and $P(\alpha,\infty)$ is the stationary P-function as evaluated in the previous section. 

\begin{figure}
 \begin{center}
  \includegraphics[width=0.47\textwidth]{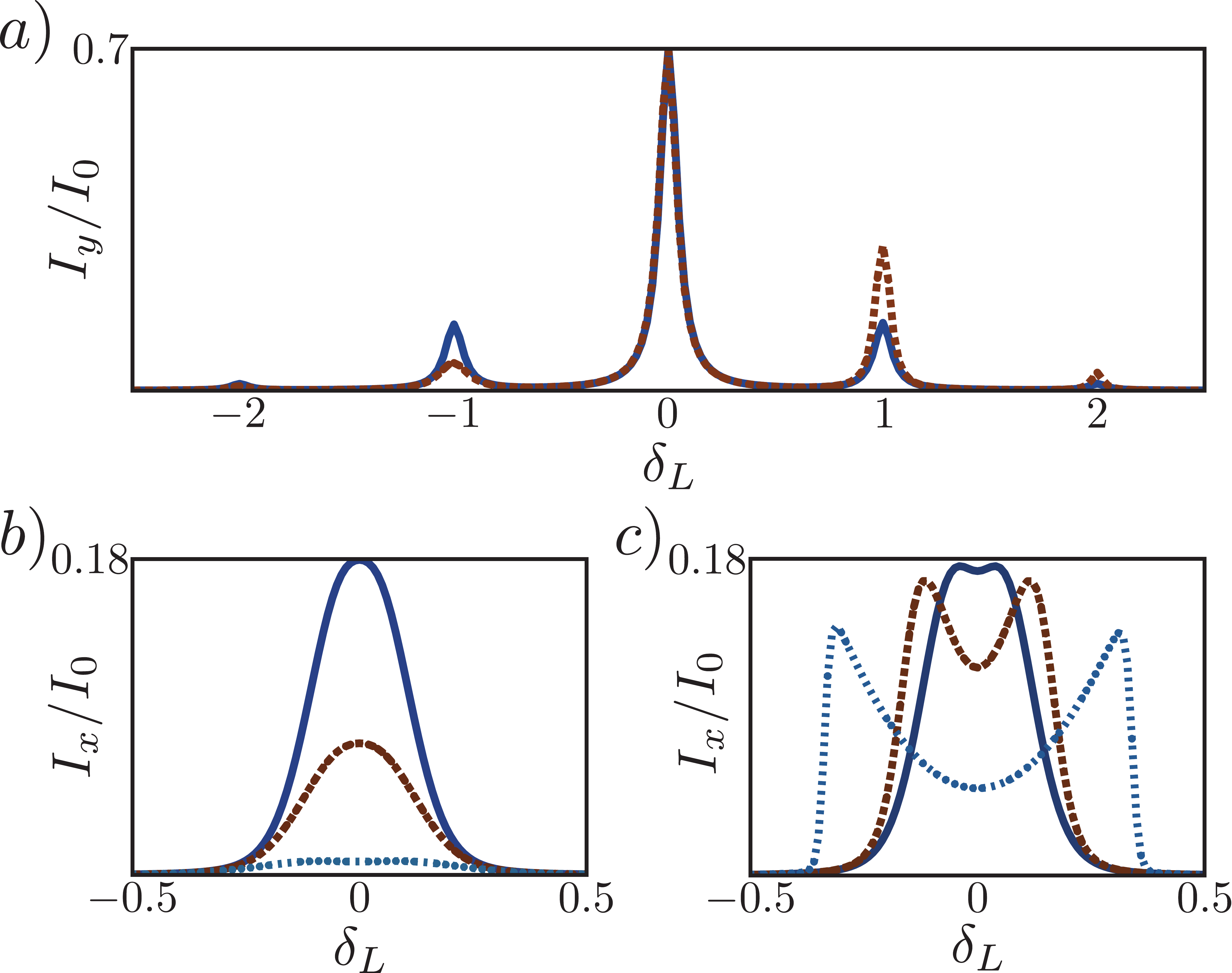}
  \caption{ (Color online). Scattered photon flux $I_{\eta=x,y}$ as functions of the laser detuning $\delta_L$ and normalized to the resonant scattering rate $I_0$. 
 a) Photon flux from the $|y\rangle$ state and assuming a dominant $\Sigma_{\parallel} $ coupling of strength $ \lambda_{\parallel} = 0.05 \omega_m$ and an equilibrium occupation number of $ N_{th} = 80 $.  At low driving, $\Omega= 0.001\omega_m$ (solid line), phonon sideband at $\delta_L=\pm\omega_m$ are of approximately the same height. At larger probe strength, $ \Omega = 0.01\omega_m$ (dashed line), the probe laser induces cooling and heating effects, which result in a pronounced asymmetry between the sidebands.   
 The other parameters for this plot are (in units of $\omega_m$) $\Gamma = 0.1$, $\Gamma_\phi=0$, $\gamma = 10^{-6}$.
 In b) and c)  the scattered photon flux from the $|x\rangle$ state is plotted for $\Delta=\omega_m$ and $\Delta=-\omega_m$, respectively. In b) the height of the scattered intensity peak provides a direct measurement of the phonon number $\langle n\rangle$. In c) the transition to the lasing regime at large $\Omega$ results in a phonon induced Rabi-splitting of the signal proportional to $\sim 2\lambda_{\perp}\sqrt{\langle n\rangle}$. For these two plots a $ \Sigma_{\perp} $-type coupling with strength $\lambda_{\perp}=0.01\omega_m$ has been assumed and $ \Omega = 10^{-2.5} $ (solid lines), $ \Omega = 10^{-2}$ (dashed lines) and  $ \Omega = 10^{-1.5} $ (dotted lines). The other parameters are as in a).  
 }\label{signal}
 \end{center}
 \end{figure}

In Fig.~\ref{signal} a) we  plot $I_y(\delta_L)$ for different driving strengths $\Omega$  
and with only $ \Sigma_{\parallel}$-type coupling.  
For clarity, we normalize each curve to $I_0=\Gamma\Omega^2/(\Gamma^2 + \Omega^2)$, which is the scattered photon flux at resonance and in the absence of the mechanical mode.  At low driving powers, the influence of the NV center on the mechanical mode is small and the resonator mode remains in a thermal state, $\langle n\rangle\approx N_{th}$. In this case we obtain the familiar phonon sideband spectrum of a two level defect \cite{HuangRhys},
\begin{equation}\begin{split}
&I_{y}(\delta_L)  \approx    \frac{\Gamma \left(\Omega/2\right)^2}{\left( \Gamma/2 \right)^2 + \delta_L^2} \sum_{n=-\infty}^{\infty} A_n e^{- \left(\lambda_{\parallel}/\omega_m \right)^2 (2 \langle n \rangle + 1)} ,
\end{split}\end{equation}
where $ A_n =  \mathcal{I}_n [2 (\lambda_{\parallel}/\omega_m)^2 \sqrt{\langle n \rangle (\langle n \rangle +1)}] \times [(\langle n \rangle+1)/\langle n \rangle]^{n/2}$ and $ \mathcal{I}_n(x) $ is the $n$th order modified Bessel function. As we increase the driving strength we find deviations from this dependence:
by probing the mechanical sidebands, we simultaneously generate significant cooling and heating, and the mean occupation $\langle n\rangle\equiv \langle n\rangle(\delta_L)$ varies as a function of the detuning. 
For example, for $\delta_L\approx -\omega_m$ the phonon modes is cooled, which leads to a reduction of the corresponding phonon peak. In the opposite case, i.e. $\delta_L\sim \omega_m$ the phonon sideband is amplified due to heating and lasing effects. The resulting asymmetry between red and the blue phonon sidebands, provides a clear signature for the backaction of the probing laser on the phonon modes.

In Fig.~\ref{signal} b) and c) we plot the scattered light intensity $I_x(\delta_L)$ from the $|x\rangle$ level, still assuming that the NV center is excited on the $|g\rangle\rightarrow |y\rangle$ transition.
In this case, there is no scattered light and $I_x(\delta_L)\approx 0$ in the absence of the mechanical mode, and therefore the measured signal is a direct consequence of phonon-induced transitions between $|y\rangle$ and $|x\rangle$.
Fig.~\ref{signal} b) shows the signal for cooling conditions, $\Delta=\omega_m$.
As above, we see that by probing the resonance with increasing driving strength, cooling sets in and reduces the height of the peak. For weak driving, $\Omega<\Gamma$ and $ \lambda\sqrt{\langle n \rangle} \ll \Gamma $, the total photon flux is approximately given by Eq.~\eqref{eq:Ix} in Sec.~\ref{sec:Idea}, 
and it can be directly used to measure the final occupation number $\langle n\rangle$. Compared to the case of a two level system described above, where the phonon sidebands are reduced by $(\lambda_{\parallel}/\omega_m)^2$, the signal given in Eq.~\eqref{eq:Ix} remains significant even for large mechanical frequencies and provides a practical way to measure the temperature of high frequency phonon modes in experiments.


Finally, Fig.~\ref{signal} c) shows the excitation spectrum $I_x(\delta_L)$ for heating conditions, $\Delta=-\omega_m$.
In this case, the transition to a lasing state can substantially increase the phonon occupation number when probing the resonance with moderate laser power. Similar to cooling, the influence of phonon lasing on the excitation spectrum can also be used to determine the mean phonon number: here, it is no longer provided by the height of the resonance, but rather the splitting of the resonance into two peaks by $\sim 2 \lambda_{\perp} \sqrt{\langle n\rangle}$. This splitting results from the mechanical system being driven into a large-amplitude oscillating state, which in turn acts like an additional strong driving field between the two excited NV states.



\section{Conclusions}\label{sec:Conclusions}
We have described the strain coupling of an NV center to an isolated vibrational mode of a diamond nanoresonator, and analyzed ground state cooling and lasing schemes for manipulating the state of that mode. In particular, we have shown that by exploiting resonant phonon transitions between two near degenerate electronic states of the NV center, cooling and lasing effects for phonons in the GHz regime can be significantly enhanced compared to similar but off-resonant effects discussed previously for two level defects. As a result, the multi-level structure of NV defects provides a versatile tool for manipulating and probing the state of individual phonon modes in nanoscale diamond structures.

\acknowledgements
We thank S. Hong, M. Aspelmeyer and Y. Chu for stimulating discussions.   
This work was supported by the EU project SIQS, the WWTF and the Austrian Science Fund (FWF) through SFB FOQUS and the START grant Y 591-N16.  
Work at Harvard is supported by NSF, CUA, DARPA, NSERC, HQOC, and the Packard Foundation.


\appendix
\section{Fluctuation Spectrum}\label{Spec}
To describe the dynamics of the NV center we use $ \sigma_{gg} = \mathbbm{1} - \sigma_{xx} - \sigma_{yy} $ and group the remaining independent expectation values into a vector,
$ 
 \langle\vec{\chi}\rangle = \left(\langle \sigma_{xx} \rangle,
\langle \sigma_{yy} \rangle,
\langle \sigma_{gx} \rangle,
\langle \sigma_{gy} \rangle,
\langle \sigma_{xg} \rangle,
\langle \sigma_{xy} \rangle,
\langle \sigma_{yg} \rangle,
\langle \sigma_{yx} \rangle
\right)^T$.
The expectation values evolve according to the Bloch equation 
\begin{equation} 
  \langle \dot{\vec{\chi}}\rangle = {\bf M} \langle\vec{\chi}\rangle + \vec{V},
\end{equation}    
where $ \vec{V} = (0,0,0,-i\Omega/2,0,0,i\Omega/2,0)^T $ and the matrix $ {\bf M} $ is explicitly given by
\begin{widetext}
 \[ { \bf M} =  \left(\begin{array}{cccccccc}
-\Gamma & 0 & 0 & 0 & 0 & 0 & 0 & 0 \\
0 & -\Gamma & 0 & i\frac{\Omega}{2} & 0 & 0 & -i\frac{\Omega}{2} & 0 \\
0 & 0 & i(\delta_L - \Delta) - \frac{\Gamma}{2} & 0 & 0 & 0 & 0 & i\frac{\Omega}{2} \\
i\frac{\Omega}{2} & i\Omega & 0 & i\delta_L - \frac{\Gamma}{2} & 0 & 0 & 0 & 0 \\
0 & 0 & 0 & 0 & -i(\delta_L - \Delta) - \frac{\Gamma}{2} & -i\frac{\Omega}{2} & 0 & 0 \\
0 & 0 & 0 & 0 & -i\frac{\Omega}{2} & i\Delta - \Gamma & 0 & 0 \\
-i\frac{\Omega}{2} & -i\Omega & 0 & 0 & 0 & 0 & -i\delta_L -\frac{\Gamma}{2} & 0 \\
0 & 0 & i\frac{\Omega}{2} & 0 & 0 & 0 & 0 & -i\Delta - \Gamma \end{array} \right).\]
\end{widetext}
 For the evaluation of the cooling rate $ \tilde{\Gamma} $ and the effective occupation number $ N_0 $, we need to calculate the spectrum $ S(\omega_m) $ given in Eq.~\eqref{spectrum}, which fully determines the cooling dynamics in the Lamb-Dicke regime. This is done using the quantum regression theorem \cite{Walls, Lambro}, and we obtain
\begin{equation}\begin{split}
S(\omega_m) = & - \left( \frac{\lambda_0+\lambda_{\parallel}}{\lambda},\frac{\lambda_0-\lambda_{\parallel}}{\lambda},0,0,0,\frac{\lambda_{\perp}}{\lambda},0,\frac{\lambda_{\perp}}{\lambda} \right) \\
& \times \frac{1}{i\omega_m \mathbbm{1} + {\bf M}} \left( \langle \vec{\chi} \bar{\Sigma} \rangle_{ss} -\langle \vec{\chi} \rangle_{ss} \langle \bar{\Sigma} \rangle_{ss} \right).
\end{split}\end{equation}
The cooling rate and the effective occupation number depend on the above spectrum as described in the main text.

\section{Fokker Planck Equation}\label{FokkerPlanck}

Starting from the set of distribution functions defined in Eq.~(\ref{distributions}), 
we use $ P_{gg} = P - P_{xx} - P_{yy} $, 
and define a vector $ \vec{P}=  ( P_{xx},P_{yy},P_{gx},P_{gy},P_{xg},P_{xy},P_{yg},P_{yx} )^T $,  which for $ \lambda \rightarrow 0 $ evolves according to
\begin{equation}
\dot{\vec{P}}(\alpha,t) = {\bf M} \vec{P}(\alpha,t) + \vec{V} P(\alpha,t) + D_{\gamma} \vec{P}(\alpha,t).
\end{equation}\\
The first two terms on the right-hand side correspond to the dissipative evolution of the NV center and ${\bf M}$ and $\vec{V}$ are defined in App.~\ref{Spec}. The third term accounts for  the mechanical damping of the oscillator, where
\begin{equation} \begin{split}
D_{\gamma} \vec{P}(\alpha,t) = & \frac{\gamma}{2} \left( \frac{\partial}{\partial \alpha} \alpha  +  \frac{\partial}{\partial \alpha^*} \alpha^* \right) \vec{P}(\alpha,t) \\
& + \gamma N_{th} \frac{\partial^2}{\partial \alpha \partial \alpha^*} \vec{P}(\alpha,t).
\end{split}\end{equation}Ê 
The coupling between the mechanical mode and the NV center is described by the term $ \dot{ \rho }(t) = -i [H_{\lambda} , \rho(t) ]$ in the master equations, where the interaction Hamiltonian is $ H_{\lambda} = \lambda \bar{\Sigma} (a e^{-i\omega_m t} + a^{\dagger} e^{i\omega_m t} )$ and 
\begin{equation}
\bar \Sigma = \frac{\lambda_{\perp}}{\lambda}\left( \sigma_{xy}+ \sigma_{yx}\right) + \frac{\lambda_0+\lambda_{\parallel}}{\lambda} \sigma_{xx}+   \frac{\lambda_0-\lambda_{\parallel}}{\lambda}\sigma_{yy}.
\end{equation} 
This coupling add the following terms to the equations of motion for the P-functions,
\begin{equation}\begin{split}
 \dot{P}_{\sigma_{jk}} = &  - i\lambda \left( \alpha e^{-i\omega_m t} + \alpha^* e^{i\omega_m t} \right) P_{[\sigma_{jk},\bar{\Sigma}]} \\
 & +  i\lambda e^{i\omega_m t} \frac{\partial}{\partial \alpha} P_{\bar{\Sigma} \times \sigma_{jk} } - i\lambda e^{-i\omega_m t} \frac{\partial}{\partial \alpha^*} P_{\sigma_{jk} \times \bar{\Sigma} },
\end{split}\end{equation}
where $P_{\sigma_{jk}}\equiv P_{jk}$. 
To remove the explicit time dependence we introduce a Floquet representation
\begin{equation}
P_{jk}(\alpha,t) = \sum_{n=-\infty}^{\infty} P_{jk}^n(\alpha,t) e^{-in \omega_m t},
\end{equation}\\
and we obtain
\begin{equation}\label{eq:dotPjk}
\begin{split}
\dot{P}_{\sigma_{jk}}^n = & i \omega_m n P_{\sigma_{jk}}^n - i\lambda \left( \alpha P_{[\sigma_{jk},\bar{\Sigma}]}^{n+1} + \alpha^*P_{[\sigma_{jk},\bar{\Sigma}]}^{n-1} \right) \\
& + i\lambda \frac{\partial}{\partial \alpha} P_{\bar{\Sigma} \times \sigma_{jk} }^{n+1} - i\lambda\frac{\partial}{\partial \alpha^*} P_{\sigma_{jk} \times \bar{\Sigma} }^{n-1}.
\end{split}\end{equation}\\
By replacing in this equation $ \sigma_{jk} $ by the identity operator $\mathbbm{1}$, we get the corresponding equation for the resonator P-function, which by including the mechanical damping, is given by
\begin{equation}\label{eq:dotPn}
\dot{P}^n = D_{\lambda} P^n + i\omega_m n P^n + i \lambda \left(  \frac{\partial}{\partial \alpha}  P^{n+1}_{\bar{\Sigma}}  -   \frac{\partial}{\partial \alpha^*} P^{n-1}_{\bar{\Sigma}}  \right).
\end{equation}\\
For the other P-distributions we obtain 
\begin{equation}\begin{split}\label{fulleq}
\dot{\vec{P}}^n = & ({\bf M} + i\omega_m n ) \vec{P^n} + \vec{V} P^n + D_{\gamma} \vec{P}^n \\ 
& + i\lambda \left(  \alpha {\bf A} \vec{P}^{n+1}  + \alpha^* {\bf A} \vec{P}^{n-1} \right) \\
& +  i\lambda \frac{\partial}{\partial \alpha} {\bf B} \vec{P}^{n+1} - i\lambda\frac{\partial}{\partial \alpha^*} {\bf B^{\dagger}} \vec{P}^{n-1},
\end{split}\end{equation}\\
where the $8\times8$ matrices $ {\bf A} $ and $ {\bf B}$  can be derived from Eq.~\eqref{eq:dotPjk}. Following Ref.~\cite{prabl} we solve this set of equations by using $\lambda \times \partial /\partial \alpha$ as a formal expansion parameter, while keeping all orders in $\lambda \alpha$. To zeroth order, and assuming $\gamma N_{th}\ll\Gamma$ the stationary solution of Eq.~\eqref{fulleq} is given by
\begin{equation}\label{eq:dotPijSS}
\left( {\bf M} + i\omega_m n \right)\vec{P}^n +  i\lambda \left(  \alpha {\bf A} \vec{P}^{n+1} + \alpha^* {\bf A} \vec{P}^{n-1} \right) = - \vec{V} P^n \delta_{n,0}.
\end{equation}
We can numerically solve this equation by truncating the maximal value of $n$ and write the result as 
\begin{equation}\label{eq:Xn}
\vec{P}^n(\alpha,t) = \vec{X}^n(\alpha) P^0(\alpha,t). 
\end{equation}
By inserting this solution back into Eq.~\eqref{eq:dotPn} we obtain
\begin{equation}
\dot{P}^0 = D_{\gamma} P^0 + i\lambda \left( \frac{\partial}{\partial \alpha} \bar{X}^{+1} - \frac{\partial}{\partial \alpha^*} \bar{X}^{-1} \right) P^0,
\end{equation}
where 
\begin{equation}
\bar X^n = \left( \frac{\lambda_0+\lambda_{\parallel}}{\lambda},\frac{\lambda_0-\lambda_{\parallel}}{\lambda},0,0,0,\frac{\lambda_{\perp}}{\lambda},0,\frac{\lambda_{\perp}}{\lambda} \right) \vec{X}^n.
\end{equation}
Now, we define parameters $ \tilde{\Gamma}(\alpha) $ and $ \Delta(\alpha) $ such that $ i\lambda \bar X^{+1} = \alpha [ \tilde{\Gamma}(\alpha) + i \Delta(\alpha) ] $. Then the above equation reads
\begin{equation}\label{eq:FPApp}
\dot{P}^0 = D_{\gamma} P^0 + \frac{1}{2}\left( \frac{\partial}{\partial \alpha} \alpha [ \tilde{\Gamma}(\alpha) + i \Delta(\alpha)] + H.c. \right) P^0.
\end{equation}
This is the result given in Eq.~\eqref{eq:FokkerPlanck}, where the small frequency shift $ \Delta(\alpha) $ has been neglected. By including in Eq.~\eqref{eq:dotPijSS} the next order correction $\sim \lambda\times \partial P^0/\partial \alpha$ we would in Eq.~\eqref{eq:FPApp} obtain additional correction to the diffusion terms~\cite{prabl}. However, a numerical estimate shows that these corrections are negligible for the high temperatures $N_{th}\gg1$ and other parameters considered in this work.

\end{document}